\documentclass[12pt] {article}
\usepackage{graphicx}
\begin{document}

\title{Fabrication of One-Dimensional Programmable-Height Nanostructures via Dynamic
	Stencil Deposition}
\maketitle

\begin{center}
J. L. Wasserman, K. Lucas, S. H. Lee, A. Ashton, C. D. Crowl, N. Markovi\'{c}\\
Johns Hopkins University, Baltimore, MD 21218
\end{center}

\begin{abstract}

Dynamic stencil deposition (DSD) techniques offer a variety of fabrication advantages not possible with traditional lithographic processing, such as the the ability to directly deposit nanostructures with programmable height profiles.  However, DSD systems have not enjoyed widespread usage due to their complexity.  We demonstrate a simple, low-profile, portable, one-dimensional nanotranslation system that facilitates access to nanoscale DSD abilities.  Furthermore we show a variety of fabricated programmable-height nanostructures, including parallel arrays of such structures, and suggest other applications that exploit the unique capabilities of DSD fabrication methods.
\end{abstract}

Fabrication of nanostructures by shadow mask deposition has been demonstrated in recent years as a means to work beyond lithographic limits. Metal can be evaporated through a nanopore in a thin membrane, leaving a nanodot as small as 10nm \cite{deshmukh, champagne}. Translation of the nanopore relative to the substrate during deposition, a technique known as dynamic stencil deposition (DSD), allows nanoscale features to be drawn directly onto the substrate. DSD has been recently demonstrated by a variety of methods such as using twin single-axis piezo actuators \cite{ono}, uncontrolled thermal motion of the mask relative to the substrate \cite{kohler}, exploiting parallax through rotation of the angle of the evaporation source relative to the mask \cite{deshmukh}, and physically translating the mask relative to the substrate \cite{racz}. More sophisticated systems have coupled the DSD mechanism to an AFM tip, allowing alignment of the deposited nanostructures with existing sample features via surface probing techniques \cite{champagne, egger}

Stencil mask deposition offers several advantages over traditional lithographic techniques. For example, employing a shadow mask stencil allows one to deposit patterned metal layers on surfaces or systems where the thermal and chemical processing of traditional lithography cannot be tolerated \cite {zhou,takano}.
 
 DSD techniques offer many additional advantages, a primary benefit being the ability to fabricate a nanostructure with a controllable height profile. Modulation of the speed of the mask's motion allows variation of the deposited feature height. Assuming a constant deposition rate, the local height of the nanostructure is proportional to the integrated time the nanopore mask uncovers it. In the static case, if a mask is held rigidly in place during deposition for an interval of time, the deposited structure will possess a height profile $M(x)$ where $x$ is the longitudinal position variable. Translating the mask's position along the $\hat{x}$ axis during deposition therefore yields a structure with a height profile given by the convolution 
\begin{equation}
h(x)=c\int t(x') M(x-x') dx'
\end{equation}
where $t(x)$ refers to the time the mask resides at longitudinal position $x$, and $c$ scales as the material deposition rate \cite{egger}.
 
DSD systems tend to be fairly complicated and difficult to design and implement, thereby inhibiting the widespread incorporation of their techniques into nanoscale research. We have constructed a compact low-cost DSD device that easily allows creation of one-dimensional programmable-height nanostructures. The device's small size and simplicity allow it to be easily adaptable to a variety of deposition chambers and systems

Our stencil masks consist of a suspended silicon nitride membrane 50nm thick with nanopore apertures\cite{kohler}. The membranes are created from Double-Side Polished $\langle 111\rangle $ Si wafers, with 50 nm of low-stress $\rm Si_{3}N_{4}$ grown via LPCVD on each side. The nitride layer is selectively removed on one side through photolithography and reactive ion etch of $\rm CF_{4}/O_{2}$. The substrate is anisotropically etched in a KOH bath, with both nitride layers acting as protective masks, leaving a suspended nitride window on the back side. Nanopores of various diameters are fabricated in this membrane, which then acts as the shadow mask.

We use two methods to create nanopores in the membrane. Pores can be milled directly through the nitride with a Focused Ion Beam (FIB). In this case we typically coat the membrane with a 10nm metallic (Cr or Pt) conductive layer to aid visualization in the SEM for alignment of the FIB. Pores as small as 50nm can be easily produced with the FIB. Likewise a layer of PMMA can be patterned with electron beam lithography, and the membrane subsequently etched in a ${\rm CF_4/O_2}$ plasma. The PMMA layer is best removed in a pure $\rm {O_2}$ plasma than with an organic solvent, to preserve the membrane and pore integrity. In either case, the nanopores produced tend to be quite larger than desired, but even a pore a few hundred nanometers in diameter can be shrunk down to a size of a few nanometers by ion-beam or electron-beam sculpting \cite{li, storm, wu, chang}.

\begin{figure}
\begin{center}
\includegraphics[width=9cm]{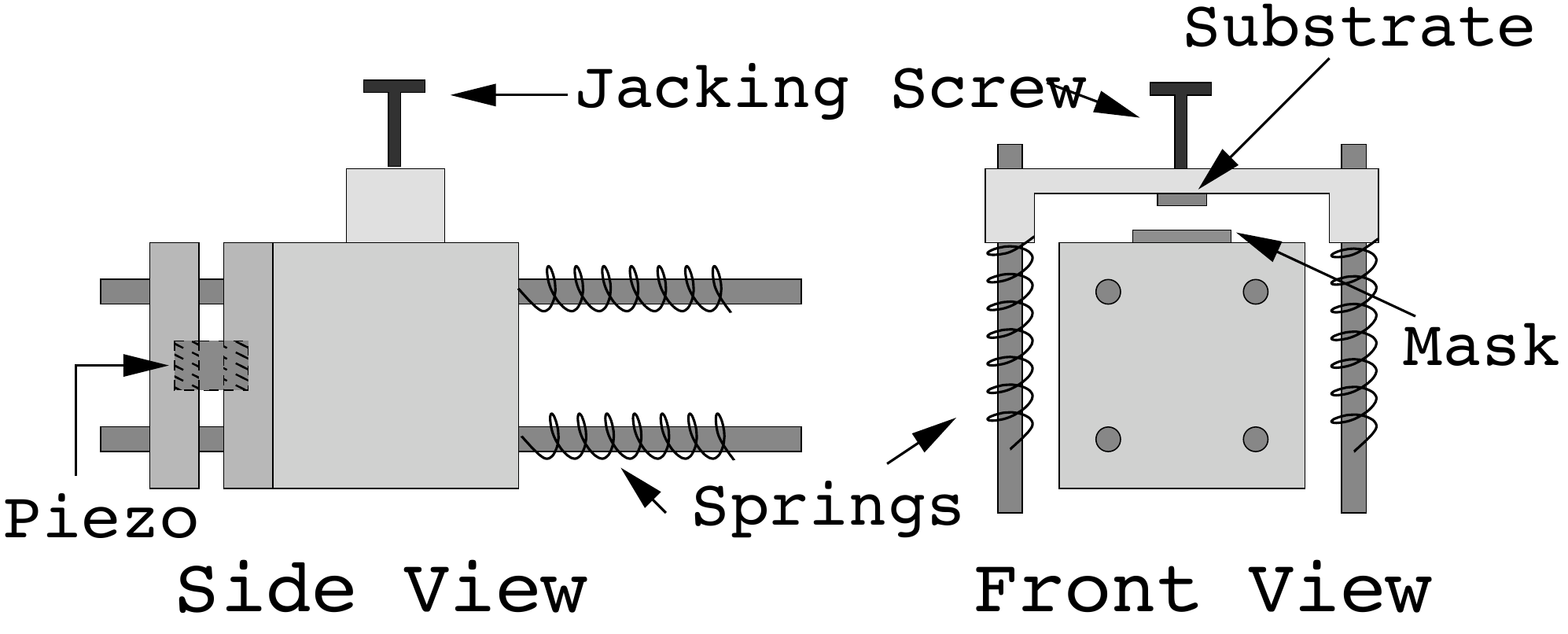}
\caption {\label{fig:assembly} Diagram of the piezo assembly for mask translation.  The rods are secured in an outer chassis (not shown), which also constrains the jacking screws.}
\end{center}
\end{figure}

The DSD nanotranslation device ultimately needs the ability to press a mask firmly against a substrate, while allowing translation of the mask as substrate remains stationary. The mask is translated relative to the substrate by means of a piezoelectric acuator, mounted in a mechanical assembly that allows translation in a single dimension. The piezo actuator is used in an open-loop fashion, for added simplicity, whereby it extends fairly linearly in a single dimension by an amount proportional to an applied DC voltage. Our actuator translates a maximum of 6 microns at 100 volts. The actuator is firmly situated between two PEEK holders, and pushes a carriage along four rods, kept under tension by four springs, as shown in Figure \ref {fig:assembly}. The substrate is mounted to a substrate holder which moves along a pair of vertical rods, also under spring-loaded tension. The spacing between the mask and substrate is controlled by three jacking screws, which keep the substrate firmly pressed against the mask during the translation process. The carriage and substrate holder rods are lubricated with a high vacuum grease to allow smooth motion. The clearance of the rods through the carriage and substrate holder is critical, there must be enough play to allow motion but enough constriction to provide the proper controllable one-dimensional constraint.

The finite solid angle of the evaporation source causes spreading of the deposited feature size beyond the size of the nanopore \cite{kohler, gross}. Geometrical considerations reveal that the size of the deposited feature is approximately given by the relation 
\begin{equation}
w_{\rm dot} \approx w_{\rm hole} + 2(d/L)w_{\rm source}
\end{equation}
The smallest possible features can be created if one minimizes the size of the evaporation source $w_{\rm source}$ and the separation gap between mask and substrate $d$. Usually $L$ is constrained physically by the size of the deposition chamber. On some occasions, for example when multiple angle evaporations of different materials are desired, it is useful to limit the separation gap $d$ to a known value by means of a spacer layer of silica spheres \cite{deshmukh}. Without using a specific spacer layer, the gap is limited to the size of any dust or contaminants residing on the substrate or mask surface, usually not much smaller than a few hundred nanometers. In order to minimize $d$, and hence our feature size, we opt not to use a spacer layer, and instead we push the mask directly against the substrate. Typically our cleanroom prevents $d$ from being less than 2 - 3$\mu $m, limiting the minimal feature sizes to about 20 to 30 nm, although better cleanroom facilities should be able to easily improve on this number.

The size of the evaporation source $w_{ \rm source}$ can be reduced to about a millimeter by employing an aperature `plug'. Gold metal is evaporated from a conical wire crucible with an oxide coating, and the molten gold forms a bead of approximately 2 - 3 mm. To reduce this evaporation size a thin tungsten circular disc about 3 mm in diameter with a 1 mm aperature is placed inside the crucible above the gold. The tungsten plug gets hot enough during the evaporation process that the gold wets the bottom plug surface and evaporates through the aperture, but does not significantly wet the top surface. Careful calibration of the plug height inside the crucible may be required to find an ideal position where the top surface remains un-wetted. This method allows efficient usage of source material, as the gold remains molten on the bottom surface of the plug. Visual inspection of the plug during deposition reveals it is fairly close to the same temperature as the crucible, by the blackbody color. Other attempts to reduce source aperture size by employing a colder aperture plate atop the crucible did work, but significant quantities of source material were wasted and the deposition rates were much slower. We expect this aperture size can be further reduced by a factor of two.

\begin{figure}
\begin{center}
\includegraphics[width=9cm]{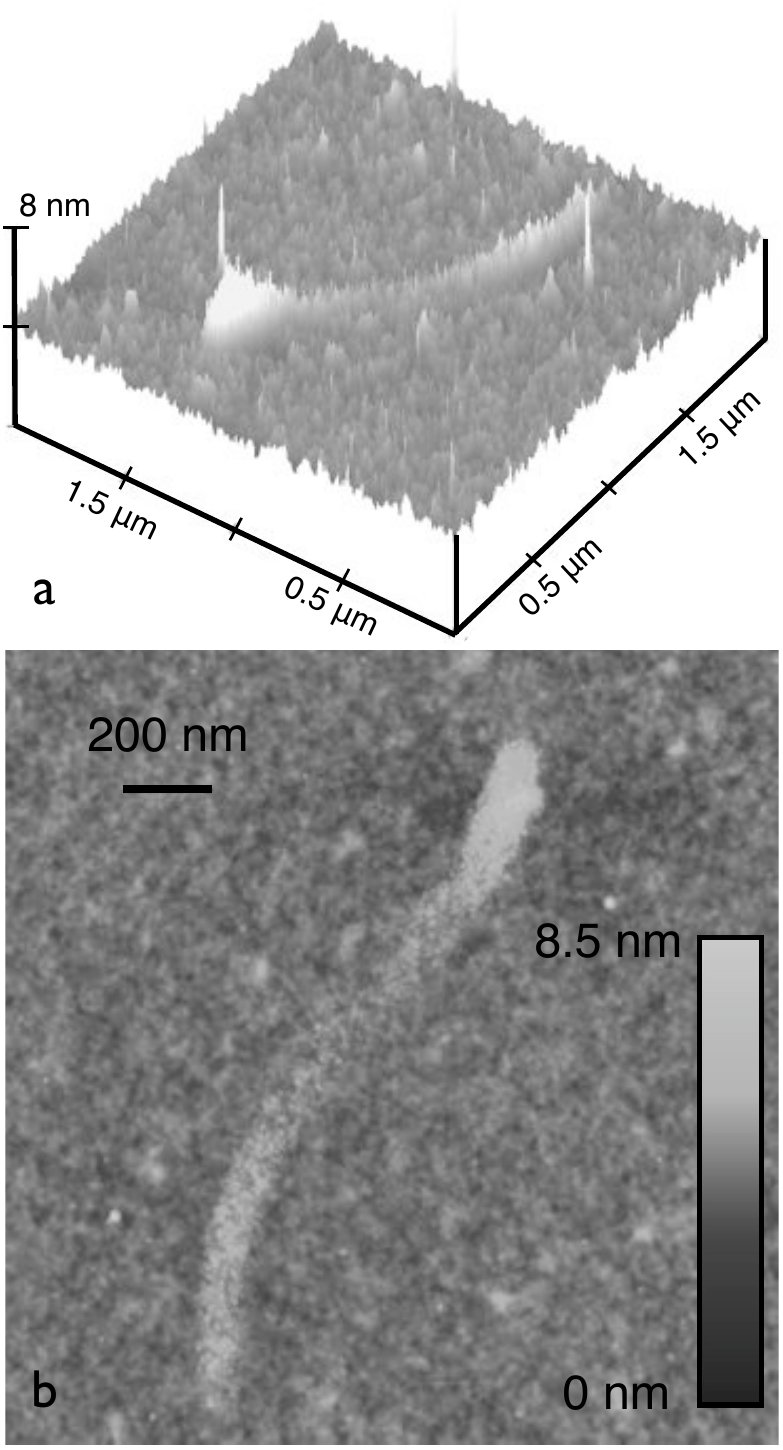}
\caption {\label{fig:wire} AFM surface plot (a) and height map (b) of a single-valley gold nanowire of width 100 nm.}
\end{center}
\end{figure}

A variety of nanostructures with programmable height profiles and feature sizes down to 50 nm have been created, and with further optimization we expect to produce features as small as 10 nm. A gold nanowire of width 100nm, with a narrow height constriction in the middle, is shown in Figure  \ref{fig:wire}, illustrating that the wire is substantially thicker at the ends than in the middle. Depending on the deposited material, such a structure could be a nanowire with a potential barrier in the middle, or an implementation of an {\em S-c-S} Josephson Junction. Controllable placement of peaks and valleys in the nanowires allows novel study of inhomogeneous nanowires as repeatable wires can be fabricated, and `grains' can be precisely positioned in a controllable fashion. All of these systems would be measurable {\em in situ}, without the need to break vacuum or perform any other processing \cite{egger2007}. Such capabilities create the opportunity to measure properties of highly-oxidizable complicated nanostructures, which could not be fabricated in any other way.

\begin{figure}
\begin{center}
\includegraphics[width=9cm]{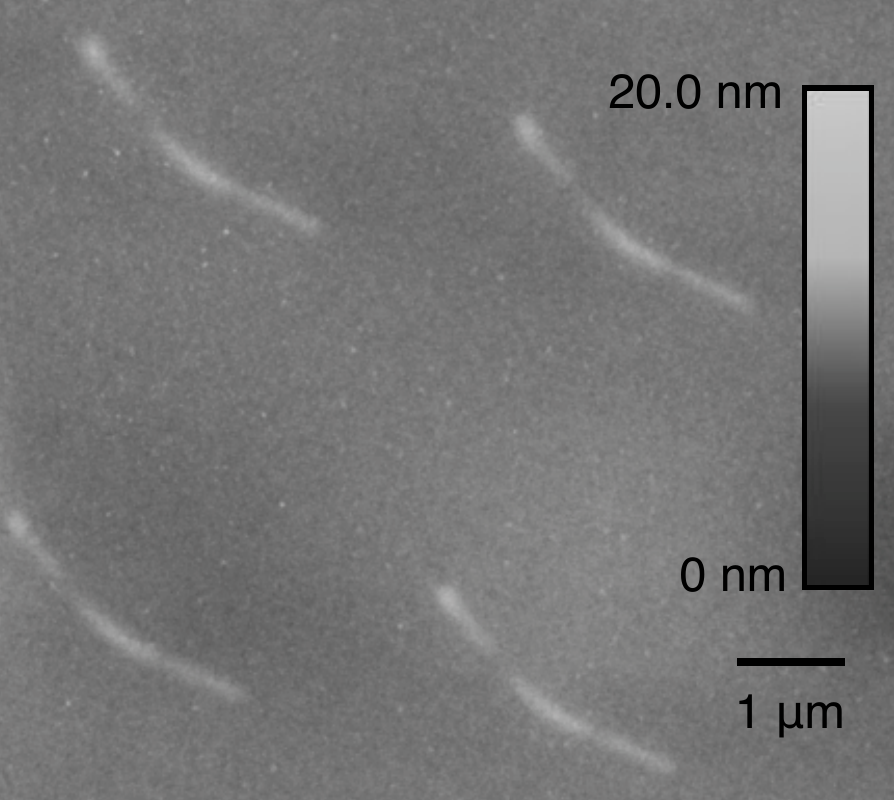}
\caption {\label{fig:array} AFM height map of a uniform parallel array of double-valley gold nanowires.}
\end{center}
\end{figure}

\begin{figure}
\begin{center}
\includegraphics[width=9cm]{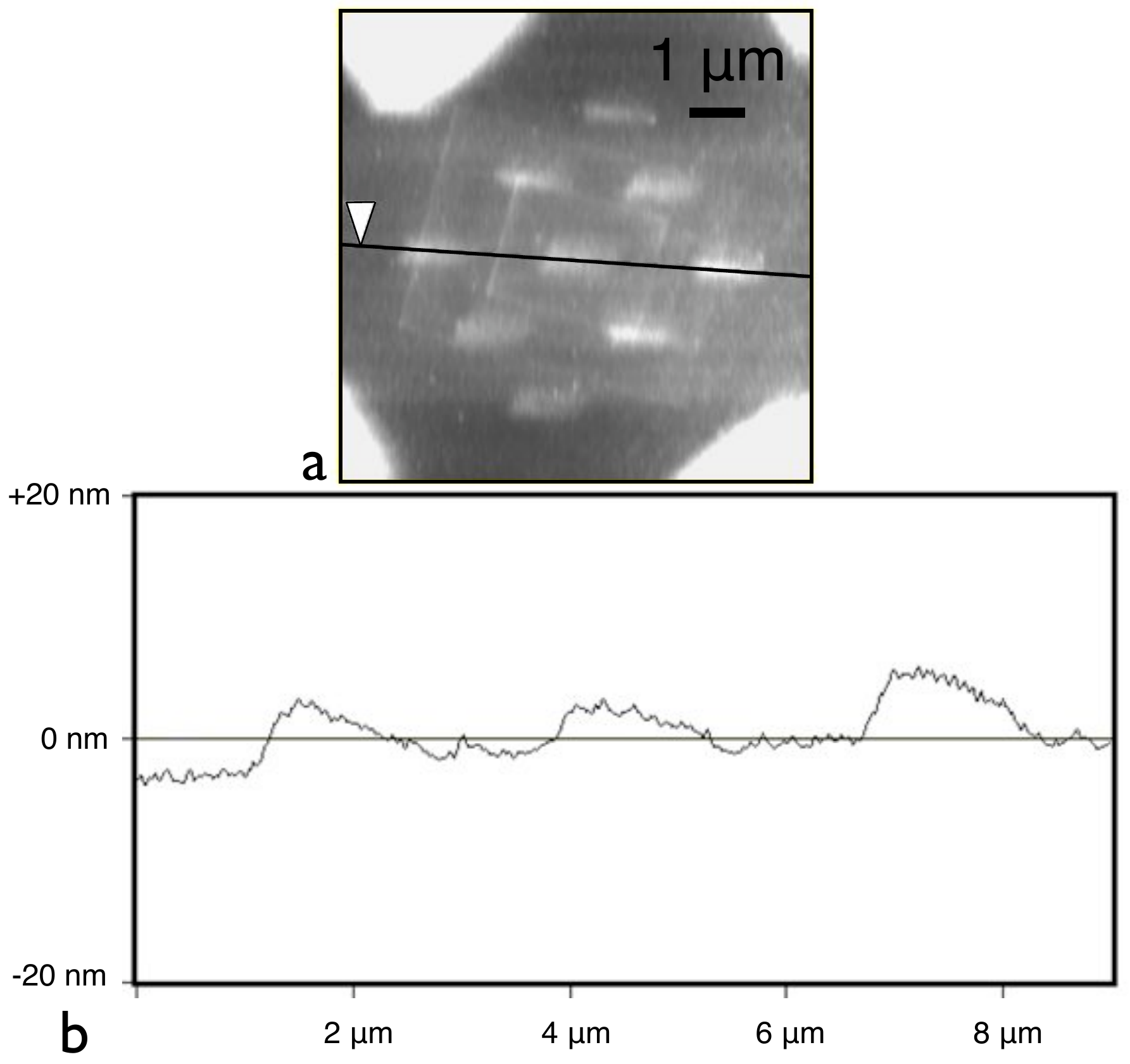}
\caption {\label{fig:nanoramps}AFM scan of an array of chromium nanoramps shown as a surface plot (a).  The cross-sectional height profile of the black line is shown in (b). Height profiles vary between nanoramps due to differing pore shapes.  The triangle in upper image is a cursor.}
\end{center}
\end{figure}

Arrays of nanostructures can be simultaneously written by using a mask with a corresponding array of nanopores. An AFM scan of an array of uniform double-valley nanowires is shown in Figure \ref{fig:array}. Each nanowire has two intermediary shallow points, and the uniformity between the nanowires in the array is quite good. A parallel array of nanoramps is shown in Figure \ref{fig:nanoramps}.a, with a cross-sectional height profile slice explicitly shown in Figure \ref{fig:nanoramps}.b. This sample used varying oblong shapes for the nanopores, which accounts for the non-uniformity between ramps.  The arrays can be over an area as wide as the mask, allowing fairly large-scale nanostructure arrays to be created. The direct ability to create identical programmable height profiles can have profound implications in the field of metamaterials \cite{shelby}. Additionally the large-scale parallel array will allow Vibrating Sample Magnetometry measurement of magnetic nanostructure systems that were not previously possible.

\begin{figure}
\begin{center}
\includegraphics[width=9cm]{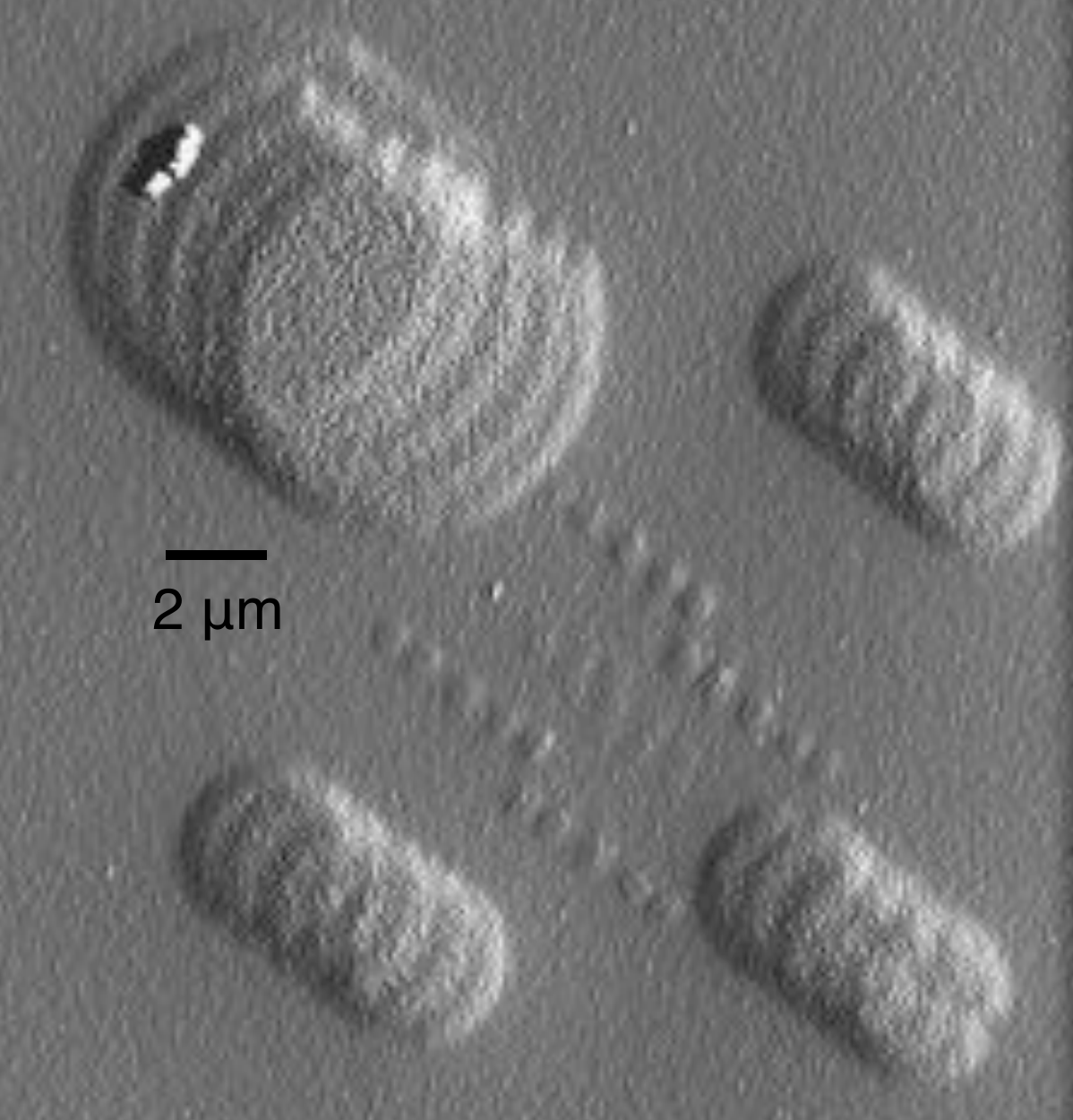}
\caption {\label{fig:circles} AFM surface image depicting overlapping gold circles of various size.}
\end{center}
\end{figure}

Interesting structures can be created by overlapping layers through discrete translation steps. Figure \ref{fig:circles} demonstrates a series of overlapping circles of various diameters, as five stationary growth periods were separated by four equal-spaced translations. The four translations were rapid enough that no appreciable material was deposited in the interim. Overlapping deposition can be used with different materials to create novel devices \cite{racz, egger2006}. However if the translation distance is larger than the nanopore size, one can create arrays of structures with irregular spacing and of varying thickness.

As with any technique, there are some limitations of the system. The motion of the piezoelectric crystal in our device is limited to one dimension, but the ability of the device to translate in a straight line depends on the mechanical rigidity of the carriage moving on the rods.  As observed in the figures, most fabricated structures do exhibit some lateral movement perpendicular to the motion of the piezo, usually no greater than 200 nm for the full range of motion. Of all nanostructures created, the largest lateral drift observed has been 400 nm over 5 $\mu$m length. We believe this is due to small amounts of slop from the rods and carriage, as well as thermal expansion. To remedy the former one can employ tighter-fitting rods, and for the latter a method for liquid cooling of the assembly to maintain constant temperature.  

For some one-dimensional applications, such lateral motion is not of great concern as the manufactured devices are still parametrically one-dimensional. If one is interested in making metallic or superconducting nanostructures, small curvature should not greatly affect the electronic boundary conditions of the device. However this limitation may be of greater concern for magnetic nanostructures.

We also envision using this nanotranslator assembly for other applications requiring carefully-controlled motion techniques. By turning the device upside-down, droplets of fluid can be placed in the pyramidal wells of the mask. This allows zeptoliter-scale quantities of fluid to be dispensed through the nanopores, onto a subsurface. Control of the dispensing rate is achieved through the same velocity modulation of the piezo actuator used for programmable height profiling. Such a dispensing mechanism has strong implications in the biotechnological sciences.

We wish to acknowledge Michael Fischbein from the University of Pennsylvania for useful discussions, Huy Vo from the Johns Hopkins MBE cleeanroom for invaluable assistance, and Steve Patterson and Scott Spangler from the Johns Hopkins machine shop for precision machining.

\clearpage

\clearpage

\clearpage

\clearpage

\clearpage

\end{document}